\documentclass[3p,times]{elsarticle} 
 
\usepackage{amssymb}
\usepackage{amsmath} 
\usepackage{amsxtra}
\usepackage{mathrsfs}
\usepackage{bm}  

\usepackage{subfig}
 
\usepackage{xcolor}

\journal{Brain Multiphysics}

\begin{document}

\begin{frontmatter} 


\title{Spatio-temporal modeling of saltatory conduction in neurons using Poisson-Nernst-Planck treatment and estimation of conduction velocity}



 
\author[add1]{Rahul Gulati}
\author[add1]{Shiva Rudraraju}
\ead{shiva.rudraraju@wisc.edu} 
\address[add1]{Department of Mechanical Engineering, University of Wisconsin-Madison, WI, USA}

\begin{abstract}
Action potential propagation along the axons and across the dendrites is the foundation of the electrical activity observed in the brain and the rest of the central nervous system. Theoretical and numerical modeling of this action potential activity has long been a key focus area of electro-chemical neuronal modeling, and over the years, electrical network models of varying complexity have been proposed. Specifically, considering the presence of nodes of Ranvier along the myelinated axon, single-cable models of the propagation of action potential have been popular. Building on these models, and considering a secondary electrical conduction pathway below the myelin sheath, the double-cable model has been proposed. Such cable theory based treatments, including the classical Hodgkin-Huxley model, single-cable model, and double-cable model have been extensively studied in the literature. But these have inherent limitations in their lack of a representation of the spatio-temporal evolution of the neuronal electro-chemistry. In contrast, a Poisson-Nernst-Planck (PNP) based electro-diffusive framework accounts for the underlying spatio-temporal ionic concentration dynamics and is a more general and comprehensive treatment. In this work, a high-fidelity implementation of the PNP model is demonstrated. This {electro-diffusive} model is shown to produce results similar to the cable theory based electrical network models, and in addition, the rich spatio-temporal evolution of the underlying ionic transport is captured. Novel to this work is the extension of PNP model to axonal geometries with multiple nodes of Ranvier, its correlation with cable theory based models, and multiple variants of the electro-diffusive model - PNP without myelin, PNP with myelin, and PNP with the myelin sheath and peri-axonal space. Further, we apply this spatio-temporal model to numerically estimate conduction velocity in a rat axon using the three model variants. Specifically, spatial saltatory conduction due to the presence of myelin sheath and the peri-axonal space is investigated.
\end{abstract}

\begin{keyword}
Action potential \sep saltatory conduction \sep signal velocity \sep Poisson-Nernst-Planck \sep electro-diffusion \sep neuronal electrophysiology \sep myelin \sep peri-axonal space \sep cable theory

\end{keyword}

\end{frontmatter}


\section{Introduction}
\label{section: introduction}

Electrical activity in nerve cells, enabled through the propagation of action potentials, is critical to the entire signaling and communication cascade of the nervous system. Disruption of this requisite signaling can lead to a number of neurological disorders such as the motor neuron diseases and is often linked with traumatic brain injury (TBI), Alzheimers, cognitive impairment, depression etc \cite{marion2018tbi, palop2007alzheimers, palop2010alzheimers, ghatak2019alzheimers, chaudhury2015depression}. To gain insight into the neuronal electrophysiology, varied experimental investigations using the patch-clamp technique, electroencephalograms (EEG), electrocardiogram (ECG), MRI, calcium imaging, voltage imaging etc have been reported in literature. Despite the wealth of information achieved by these investigations, there is a need to supplement these studies with a robust numerical implementation which has the potential to represent the electrophysiology to a far greater resolution as recorded by the experiments.

Based on numerous voltage clamp experiments on the giant squid, Hodgkin-Huxley came up with a first mathematical model in the form of an electrical circuit to describe the current through the neuronal membrane \cite{hodgkin1952huxley}. They quantitatively detailed the respective ionic conductance in respect to the membrane voltage. Huxley delineated that the action potential propagation along the axon closely follows Ohm's law \cite{huxley1959ohms}. Using cable theory, they arrived at a one dimensional model of the action potential propagation. Based on the physiology of the neuron, the importance of myelin sheath that surrounds the axon, has been emphasized. Degradation of this protective covering for example with age can lead to slowdown of the signal or even signal disruption resulting in various diseases \cite{duncan2016myelindiso, zhan2014myelindiso2}. To be able to study the effect of myelin on action potential propagation, the Hodgkin-Huxley model was modified to incorporate the myelin sheath and to obtain the single-cable model \cite{cohen2020cell, richardson2000}. Experimental investigation of the action potential by Barrett and Blight in the 1980's lead to the discovery of an after-potential \cite{barett1981barrett, blight1985, blight1985someya}. With the advent of advanced microscopy techniques, the existence of a secondary electrical conduction pathway in the peri-axonal space under the myelin sheath has been uncovered \cite{cohen2020cell}.

The cable theory based models namely Hodgkin-Huxley, single-cable, double-cable, etc., have greatly contributed to our understanding of the neuronal electrophysiology. They provide us an excellent insight into the membrane potential, action potential propagation along the neuron, the electric current propagating along the axon, the effect of saltatory conduction due to the presence of myelin sheath, the effect of the submyelin peri-axonal space dictated by the double-cable model, the conduction velocity implied by each of these models etc. The cable theory based models, however, have some limitations. {First, these are a one dimensional reduction of the complex heterogeneous spatial propagation of the action potential. Secondly, the cable theory based models fail to describe the underlying spatial ionic diffusion and the generated electric field during the electrical conduction propagation. Therefore, these models are not able to accurately describe the dynamics after a prolonged electrical activity or when the diameter is relatively smaller, as in the case of dendrites. Further, the cable theory based models cannot be easily extended to account for the membrane microenvironment, such as incorporating the membrane-glia interaction.} 
 
The Poisson-Nernst-Planck based electro-diffusive model has the capability to overcome the limitations of the cable theory based models and provide a spatio-temporal representation of the electrical potential along with the ionic distributions \cite{pods2013pnp}. The PNP model is a more generalized model which can be reduced to the electroneutral model and can subsequently be reduced to the one dimensional cable theory based model \cite{mori2009PNP2Cable}. Qian and Sejnowski modelled one of the first intracellular dynamics incorporating one dimensional PNP theory \cite{qian1999pnp}. PNP model has also been applied to neuron-ECM-astrocyte interations \cite{halnes2013astrocyte}. Assuming electroneutrality, ionic dynamics have also been represented using Kirchoff-Nernst-Planck \cite{mori2009electroneutrality, halnes2016electroneutrality}. However, the assumption of electroneutrality for nonuniform geometries is invalid \cite{lopreore2008pnp2}. Using PNP model, the neuronal  intracellular-extracellular dynamics have been represented for a single node of Ranvier \cite{lopreore2008pnp2, dione2016pnp1}. It has been demonstrated that the dynamics of the PNP model resemble to that of the cable theory at higher ion channel density \cite{lopreore2008pnp2}. The PNP model can also be coupled with mechanics to represent the complex neuronal mechano-electrophysiology interactions \cite{goriely2015brain, garcia2019mechanicsElectro, kwong2019mechanicsElectro2,deb2021shellMechanics}.

In this work, we extend the electro-diffusive PNP model to multiple nodes of Ranvier, enhancing our ability to study the electrophysiology in a full length neuronal axon. We present novel variants of the PNP model based on the discrete cable theory based models, i.e. PNP model, PNP model with myelination, and PNP model with myelin and peri-axonal space. {As an example, we demonstrate these models by simulating action potential conduction in a rat neuron}. Spatial saltatory conduction due to the presence of myelin sheath and the peri-axonal space is demonstrated. Finally, we provide a detailed insight into the numerically estimated conduction velocity for a rat and squid neuron using various representative PNP electro-diffusive models. The Finite element (FE) method is used to discretize the set of PDEs underlying the PNP model. As suggested by an earlier work, non-homogeneous adaptive mesh is employed \cite{dione2016pnp1}. Results indicate that the conduction velocity (CV) of the action potential increases with the presence of myelin sheath and the peri-axonal space. The CV for the PNP with myelin is comparable to the single cable network but the CV for the PNP with myelin and peri-axonal space model does not increase drastically as compared to the double cable model. We observe that the action potential amplitude is lower for the PNP model when the myelin sheath is present.

In section \ref{section: electricalNetwork1D}, we briefly review and illustrate the well known models based on the one dimensional cable theory. The electro-diffusive PNP model is presented in section \ref{section: PNP}. The mathematical formulation of the numerical framework is elaborated in section \ref{mathematical formulation}. The simulation results of the various models of the PNP are detailed in section \ref{section: results}. Finally, a discussion of the conduction velocity for a rat and a squid neuron is in section \ref{section:discussion}, followed by conclusion in section \ref{section: conclusion}.

\section{Review of electrical network models of action potential propagation}
\label{section: electricalNetwork1D}

\begin{figure}[h]
  \centering
\includegraphics[width=1.0\linewidth]{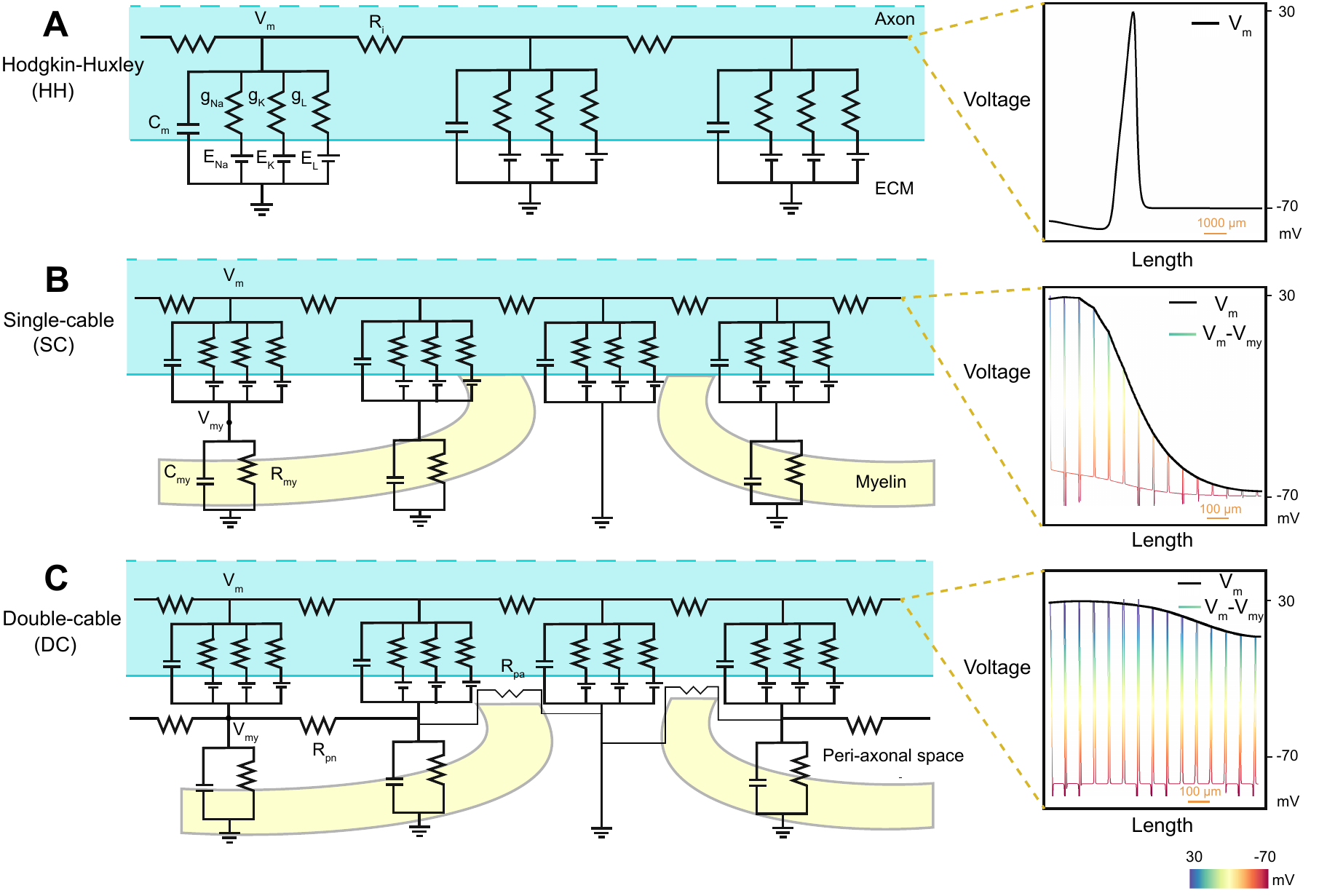}
\caption{Electrical circuit of cable theory based models. (A) Schematics of cable theory based Hodgkin-Huxley circuit consists of membrane capacitance, resistance offered by the ion channels. Action potential propagates like a soliton through the axon and is depicted on the right. Extracellular voltage is taken as 0 mV. (B) Electrical circuit of the single-cable model considers  the presence of myelin sheath. The capacitance and resistance offered by the myelin are explicitly modeled. (C) Electrical network of the double-cable representation incorporates an additional cable pathway for the submyelin peri-axonal space. Action potential jumps from one node of Ranvier to the next, resulting in a higher conduction velocity.}
 \label{fig:comparisonCableModels}
\end{figure}

\subsection{Hodgkin-Huxley model}
The classical work of Hodgkin and Huxley was a landmark model in terms of providing deep insights into the ionic basis of action potential propagation in nerve cells. Based on the voltage clamp experiments on the Giant Squid, the physiology of the initiation and propagation of action potential in a neuron was posed as a coupled set of ordinary differential equations. This electrical network model takes into account the membrane capacitance and the ionic currents due to the sodium ions, potassium ions and some leak current through the respective ion channels in the neuronal membrane. The influence of the conductance of the respective ion channels, or conversely the resistance, on the action potential was quantitatively estimated using experimental data. The resulting electrical circuit is depicted in Figure \ref{fig:comparisonCableModels}(A), and the corresponding governing equation linking the evolution of the membrane potential with the underlying ionic transport is the following:
\begin{equation}
    C_m \frac{\partial V_m}{\partial t} + \bar{G}_{Na} m^3 h (V_m-E_{Na}) + \bar{G}_K n^4 (V_m-E_K) + G_L(V_m-V_{rest}) = I_{inj}
\end{equation}
where $C_m$ is the membrane capacitance of the lipid bilayer, $V_m$ is the membrane potential, $V_{rest}$ is the resting potential of the nerve cells and $I_{inj}$ is the injected current through the voltage clamp experiments to initiate the action potential in the cell. $\bar{G_i}$ and $E_i$ are the peak conductance and the Nernst potential of the  sodium or potassium ions, and ${G_L}$ is the conductance of the membrane leak channel.

Numerous experiments have pointed towards the effective behaviour of the intra-cellular region to that of a resistor along the axon \cite{huxley1959resistance}. Utilizing cable theory, one can then arrive at:
\begin{equation}
    \frac{1}{R_i} \frac{\partial^2 V_m}{\partial x^2} = I_{inj}
\end{equation}
where $R_i$ is the resistance per unit length of the intra-cellular region along the axon. The partial differential equation (PDE)  form of the Hodgkin-Huxley equation can then be written as:
\begin{equation}
    C_m \frac{\partial V_m}{\partial t} + \bar{G}_{Na} m^3 h (V_m-E_{Na}) + \bar{G}_K n^4 (V_m-E_K) + G_m(V_m-V_{rest}) = \frac{1}{R_i} \frac{\partial^2 V_m}{\partial x^2} \label{eq:HH}
\end{equation}
Using the above one-dimensional PDE, one can model the propagation of action potential along the length of the axon.

\subsection{Single-cable model}
From the physiology of nerve cells, it is well known that the glial cells provide a protective covering to the axonal membrane. This protective covering consists of multiple layers of myelin sheath along the axon, with gaps in between. These axonal gaps, void of any myelin covering, are identified as the nodes of Ranvier. These myelin lamellae play an indispensable role in the rapid movement of the action potential, since the direct ionic exchange with the extra-cellular medium occurs only at the nodes of Ranvier. This leads to a local current and the action potential jumping from one node of Ranvier to the next, more commonly referred to as {\em saltatory conduction}. The degradation of the myelin layers is known to lead to a decline in the conduction velocity and is also linked with various neuronal disease conditions \cite{duncan2016myelindiso, zhan2014myelindiso2}.

To accommodate this spatial heterogeneity of the myelin sheath, the standard Hodgkin-Huxley treatment needs to be altered to model myelinated-axons. A simple electrical circuit to realize this, known as the single-cable model, has been proposed \cite{richardson2000, koch2004}. The ionic exchange between the intra-cellular regions and the extra-cellular regions takes place at the membrane, but only at the location of the nodes of Ranvier. As an example, for a rat axon, the span of the nodes of Ranvier is around $2.3 \mu m$ and the nodes are separated by distance of $ 70 \mu m - 100 \mu m$  \cite{cohen2020cell}. The single-cable model with the myelin sheaths having their respective capacitance and resistance is shown in Figure \ref{fig:comparisonCableModels}(B) \cite{cohen2020cell}.

\subsection{Double-cable model}
The presence of the submyelin peri-axonal region has been proposed as a potential pathway for rapid electrical conduction along the axon \cite{cohen2020cell, stephanova1995bostock}. Due to the presence of two conduction pathways, each modeled using cable theory, this treatment is referred to as the double-cable model.  An electrical circuit representing such a double-cable model can be seen in Figure \ref{fig:comparisonCableModels}(C) .
The basic governing equations for the electrical circuit at the node of Ranvier are similar to the set of equations from the single cable model/Hodgkin-Huxley model. In addition, using cable theory, the partial differential equation modeling the peri-axonal space takes the form:
\begin{equation}
    \frac{1}{R_i} \frac{\partial^2 V_m}{\partial x^2}  + \frac{1}{R_{pa}} \frac{\partial^2 V_{my}}{\partial x^2} = C_{my} \frac{\partial V_{my}}{\partial t} + \frac{V_{my}}{R_{my}}
\end{equation}
where $R_{pa}$ is the resistance per unit length in the peri-axonal space, $C_{my}$ is the cumulative capacitance of the myelin sheath, $R_{my}$ is the myelin resistance and $V_{my}$ is the potential in the peri-axonal region.

\subsection{Comparison of action potential profiles for electrical network models}
The main results of this manuscript will be discussed in Section~\ref{section: results}, but it is worthwhile to present here a brief comparison of the relative differences between the action potential (voltage) profiles predicted by modeling the electrical network models described above. Figure \ref{fig:comparisonCableModels} presents such a comparison of the voltage profiles. All these plots were generated by solving the governing equations listed earlier in this section using an in-house 1D Finite Element Method implementation, and the model input parameters available in the literature for a giant squid axon \cite{hodgkin1952huxley} and rat axon \cite{cohen2020cell} were used.

Shown as part of Figure \ref{fig:comparisonCableModels}(A) is a typical voltage profile predicted by the Hodgkin-Huxley model. The voltage profile is a single spike that propagates along the length of the axon, in this case from left to right.  After the refractory period of the propagating spike, if a potential/current perturbation exceeding the threshold value is injected at the left boundary, then it would result in a voltage wave traveling with the same amplitude.

The voltage profile produced by a single-cable model is shown in Figure \ref{fig:comparisonCableModels}(B). As can be seen, the potential difference between the membrane potential and the myelin potential, ($\text{V}_m-\text{V}_{my}$) jumps across the nodes of Ranvier. This saltatory conduction leads to fast propagation of the voltage envelope, i.e. the membrane potential, which travels like a soliton-like wave. The decreased capacitance and increased resistance due to the myelin sheath enables this mode of propagation. In this simulation, the nodes of Ranvier are assumed to be equidistant. However, if the inter-nodal distance is varying along the length of an axon, it will lead to an axially varying conduction speed.

The voltage profile produced by a double-cable model, that incorporates a secondary conduction pathway, is shown in Figure \ref{fig:comparisonCableModels}(C). It can be observed that the profile of the voltage envelope is flatter than the voltage profiles predicted by the previous two models. Like in the case of the single-cable model, the action potential jumps from one node of Ranvier to the next, but through the peri-axonal space, and the propagation is relatively much faster. The peak amplitude is comparable to the peak amplitude for the Hodgkin-Huxley model and single-cable model.
\section{A field theoretic model of action potential propagation}
\label{section: PNP}

Cable theory based electrical network models have proven to be fundamental to our current understanding of the dynamics involved in action potential propagation. However,
at the core, the network models are reduced order representations that try to capture the complex spatio-temporal variations of the ionic transport, voltage distribution, and most importantly the membrane structural heterogeneity into effective electrical properties like capacitance and resistance of the membrane and the channels. As mentioned in the introduction, a deeper investigation into spatial and temporal interactions of the ionic transport with the membrane microstructure (ion channel and pump distributions, myelin, glial environment, etc.) is desired, but the resulting coupled evolution of the voltage distributions becomes necessary when more complex phenomena like neuronal injury effects on action potential propagation \cite{bar2016strain, li2011effects, estrada2021neural}, and conditions like neuronal hyperexcitability observed with Alzheimer's disease \cite{palop2016network, ghatak2019mechanisms, kim2007bace1} are to be modeled. \\
An electro-diffusive framework for modeling spatio-temporal ionic charge distribution and the resulting voltage evolution using high-fidelity partial differential equations (PDE)  modeling coupled electrostatics and electrochemistry can be capable of faithfully representing spatio-temporally heterogenous evolution of the ionic and voltage distributions leading to generation, propagation and potentially disruption of the neuronal action potential. While not addressed in this current manuscript, such a capability to model spatial heterogeneity and membrane geometry-action potential interactions is evidently more important in the dendrites than the axons, due to the complex morphology of neuronal dendritic structures and synapses. {We now propose a PDE based field theoretic implementation of the Poisson-Nerst-Planck (PNP) framework that can model 2D/3D ionic and voltage field distributions and their interactions with the membrane microstructure.} \\
In this model, the transport of the respective ionic species, $c_i$, due to the corresponding diffusive and electromigration flux, $\boldsymbol{F}_i$, is modeled using the classical Nernst-Planck equation:

\begin{equation}
     \frac{\partial c_i}{\partial t} = - \nabla \cdot \boldsymbol{F}_i \label{eq:PNPConc}
\end{equation}

The ionic flux comprises of the diffusion term and the electro-migration term as presented in Eq. \ref{eq:PNPFlux}. Here $D_i$ is the diffusion coefficient of the $i^{th}$ ion, $R$ is the gas constant, $T$ is the temperature, $F$ is the faraday constant and $z_i$ is the valency of the $i^{th}$ ionic species.  It is to be noted that the cable theory based models only account for the electro-migration term i.e. the voltage gradient term. While the diffusion term may be neglected for axons with larger diameter, for finer geometries like dendrites, the diffusion term takes over the electro-migration term \cite{qian1999pnp}. Hence, the PNP treatment has a higher-fidelity in representing the underlying action potential propagation. The total flux is given by:

\begin{equation}
    \boldsymbol{F}_i = -[D_i (\nabla c_i + \frac{c_i F z_i}{RT} \nabla V )] \label{eq:PNPFlux}
\end{equation}

The coupling between the voltage and local ionic distribution is modeled with Poisson's equation \cite{pnp2008gauss2pnp}, shown below.
\begin{equation}
    -\nabla \cdot (\epsilon \nabla V) = F \sum_{i=1}z_i c_i \label{eq:poisson}
\end{equation}
where $\epsilon$ is the permittivity of the medium. In this work, we consider three ionic species: $Na^+$, $K^+$ and $Cl^-$. The presence of an anion species ensures regulation of the required potential - for instance at the resting state. The sodium and potassium ions, as is well understood, depolarise and re-polarise the neuron. While the above PDE formulation by itself is well known, the novelty of its implementation in this work is from its application to model field variations and interactions in the presence of geometric heterogeneity of the nodes of Ranvier, myelin distribution, and the peri-axonal space.

\subsection*{Cable theory models as special cases of the general PNP framework}
\begin{figure}[h!]
  \centering
\includegraphics[width=1.0\linewidth]{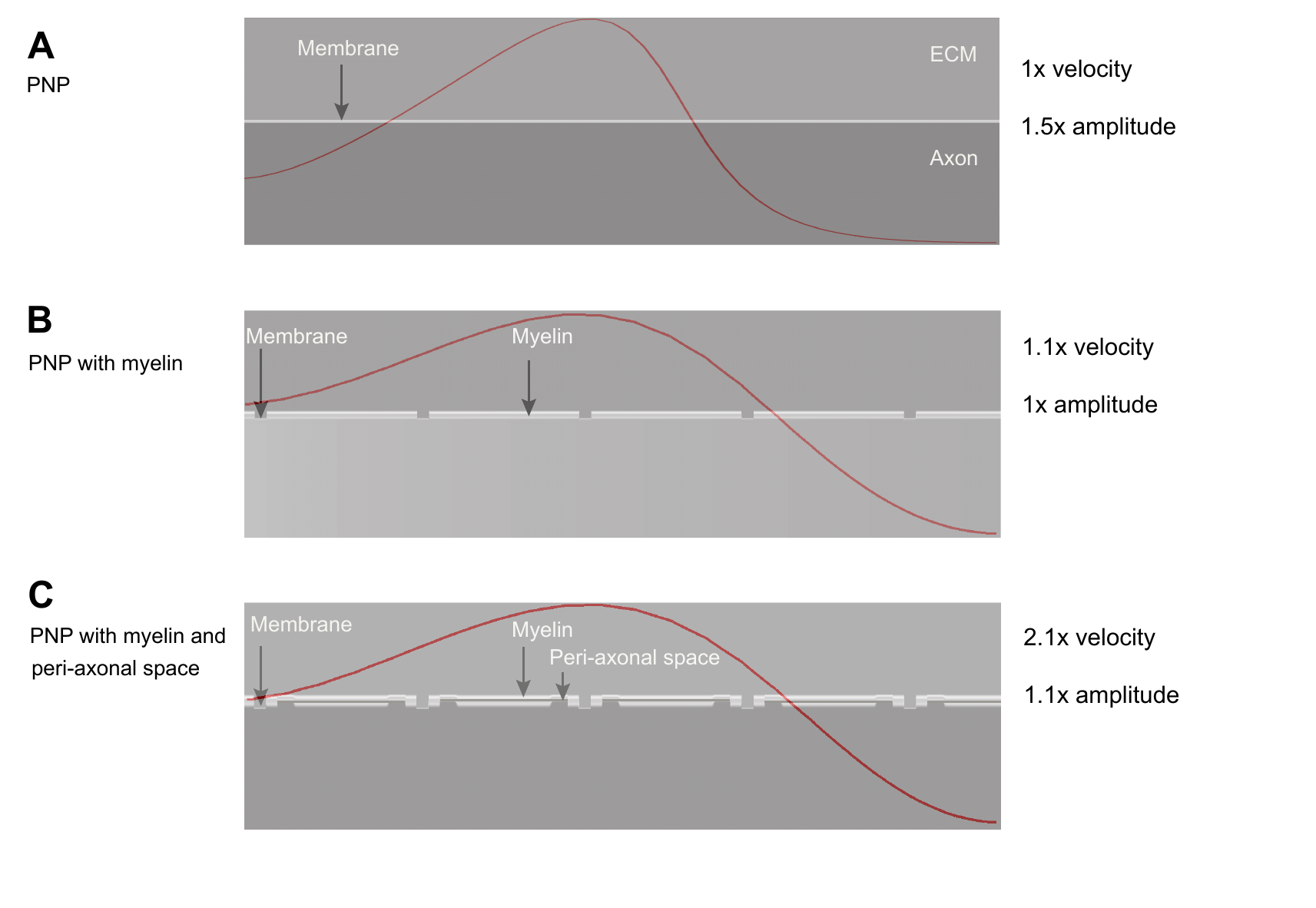}
\caption{Schematic of adaptations of the PNP model based on various cable theory based models. Normalised profile of the action potential propagation computed in this work using the corresponding PNP model is depicted for the three variants: A) The PNP model comprises of ionic exchange between the intra-cellular and the extra-cellular region through the ionic channels present uniformly on the neuronal membrane. B) PNP with myelin model unifies the PNP model and the presence of myelin sheath. The ion channels are only present at the nodes of Ranvier. The low capacitance of the myelin sheath steers the action potential with higher velocity. C) The presence of the submyelin peri-axonal space is accounted for in the PNP model with myelin and submyelin peri-axonal space.}
 \label{fig:spatialModelVariations}
\end{figure}

As advancements in high resolution imaging techniques lead to an improved understanding of the neuronal membrane microenvironment, various electrical network models incorporating the membrane spatial heterogeneity have been proposed over the years. Of these models, a primary classification in terms of increasing complexity of the membrane heterogeneity leads to the three treatments described above: Hodgkin-Huxley model, single-cable model and double-cable model. The PNP model is capable of representing the spatial heterogeneity of each of these models, and thus reproduce the action potential conduction profiles observed with these electrical network models. {\em To the best of our knowledge, numerical demonstration of this equivalence for each of these three models, and the extraction of equivalent 1D electrical network results (effective capacitance, effective resistance, action potential profile, etc.) from a more general electro-diffusion framework, has not been shown earlier in the literature.}
Figure \ref{fig:spatialModelVariations} depicts this equivalence between:
\begin{itemize}
\item PNP model and the classical Hodgkin-Huxley model,
\item PNP model with myelination and the single-cable model,
\item PNP model with myelin plus peri-axonal space and the double-cable model.
\end{itemize}

{The axonal membrane has ion channels and the embedded membrane capacitance in the PNP model similar to the Hodgkin-Huxley model. The existence of the myelin sheath leads to the presence of nodes of Ranvier and decreased internodal capacitance in the PNP model with myelin. Note that there is no direct correlation of $V_{my}$ as in the single cable model. Finally, the PNP model with myelin and peri-axonal space incorporates the sub-myelin peri-axonal space, leading to lower net resistance and faster propagation speed of the action potential. Here, the axial resistance and the peri-axonal resistance act in parallel, and hence the net axial resistance is lower. While the peri-axonal resistance does not correlate to the axial resistance in the double cable model, this is not the case for the equivalent PNP model as the resistance is computed using the underlying ionic constants. Here, the diffusion constants of the underlying ions in the peri-axonal space are assumed to be the same as in the cytoskeleton region.}

{Figure \ref{fig:spatialModelVariations} also shows, as an overlaid line plot in each case, the 1D action potential profile (obtained as a line-out along the axonal axis from the 2D PNP simulations) for the three PNP model variants listed above. It can be noticed that the curvature of the voltage potential is lower when myelin is present as it leads to faster action potential propagation. Further, the presence of myelin leads to a lower action potential amplitude. This maybe due to the fact that the action potential jumps node to node as soon the threshold potential is attained.}

\section{Numerical implementation of the PNP model}
\label{mathematical formulation}
The coupled nonlinear system of PDE's for ionic concentration and electrical potential in the electro-diffusive PNP model are solved using the standard Finite Element Method (FEM). The primal fields that are solved for are the voltage and the concentration of $Na^+$/ $K^+$/ $Cl^-$ ions. The electric field and the net charge are derived fields. The salient features of the computational implementation are: adaptive mesh refinement near the nodes of Ranvier, adaptive time-stepping schemes, support for parallel direct and iterative (Krylov-subspace) solvers with Jacobi/SOR preconditioning. The weak formulation of the governing equations solved with FEM are given in Section \ref{sec:feModelling}. The computational framework is made available to the wider research community as an open source library~\cite{GitRepo2022}, and we hope it serves as a platform for wider adoption of the high-fidelity PNP framework by neuronal modeling researchers.

\subsection{Weak formulation of the PNP model}
\label{sec:feModelling}
The Nernst-Planck and Poisson equations, expressed in their weak (integral) formulation that is suitable for the FEM implementation, following standard notation, are as follows:

\vspace{0.2in} 

\noindent Find the primal fields $\{ V, c_{Na}, c_{K}, c_{Cl} \}$, where,
\begin{align*}
V &\in \mathscr{S}_{V},  \quad \mathscr{S}_{V} = \{ V  ~\vert V   = ~\bar{V} ~\forall ~\textbf{X} \in \Gamma^{V}_g \}, \\
c_{i} &\in \mathscr{S}_{c_i},  \quad \mathscr{S}_{c_i} = \{ c_i  ~\vert c_i   = ~\bar{c}_i ~\forall ~\textbf{X} \in \Gamma^{c_i}_g \}
\end{align*}
and $i \in \{Na^+, K^+, Cl^-\}$, such that,   
\begin{align*}
\forall ~w^{V} &\in \mathscr{V}_{V},  \quad \mathscr{V}_{V} = \{ V  ~\vert V   = ~0 ~\forall ~\textbf{X} \in \Gamma^{V}_g \}, \\
\forall  ~w^{c_i} &\in \mathscr{V}_{c_i},  \quad \mathscr{V}_{c_i} = \{ c_i  ~\vert c_i   = ~0 ~\forall ~\textbf{X} \in \Gamma^{c_i}_g \}
\end{align*}
we have,
\begin{equation}
    -\frac{F}{\epsilon} \int_{\Omega} w^V~(c_{Na}+c_{K}-c_{Cl}) ~dV  +\int_{\Omega} \nabla w^V  \cdot \nabla V ~dV - \int_{\Gamma^{V}_h} w^V~(\nabla V \cdot \boldsymbol{n}) ~dS =0 \label{eq:weakPoisson}
\end{equation}
and,
\begin{equation}
    \int_{\Omega} w^{c_i}~\frac{\partial c_{i}}{\partial t} ~dV  + \int_{\Omega} \nabla w^{c_i} \cdot ~D_i \nabla c_i ~dV + \int_{\Omega} \nabla w^{c_i} \cdot ~D_i \frac{c_i z_i F}{R T} \nabla V ~dV  + \int_{\Gamma^{V}_h} w^{c_i} ~(\boldsymbol{F}_i \cdot \boldsymbol{n}) ~dS = 0 \label{eq:weakNP}
\end{equation}
where $w^{V}$ is the variation for the voltage field, $w^{c_i}$ are the variations for the ionic fields. $\Omega$ is the problem geometry, $\{ \Gamma^{V}_g, \Gamma^{c_i}_g \}$ are the Dirichlet boundaries and $\{ \Gamma^{V}_h, \Gamma^{c_i}_h \}$ are the Neumann (Flux) boundaries of the voltage and ionic fields, respectively. $\boldsymbol{n}$ is the unit normal vector . In this work, there is no voltage flux ($\nabla V \cdot \boldsymbol{n} =0$) at all the boundaries. Eq. \ref{eq:weakNP} and Eq. \ref{eq:weakPoisson} are the governing equations solved using FEM. \\

The PNP framework models the evolution of the voltage field and the ionic concentrations of $Na^+$, $K^+$ and $Cl^-$. The initial ionic concentrations in the various regions such as the extra-cellular region, membrane, myelin, cytoskeleton, etc., are mentioned in the Supplemental Information. The initial voltage in the ECM is taken to be 0 mV and in the cytoskeleton region to be the resting value of -70 mV. The boundary conditions on the top and bottom surface of the domain are applied so that the fields, i.e. voltage and the ionic concentrations, represent their bulk value in the extra-cellular region as in Eq. \ref{eq:dbc}. The ionic exchange at the nodes of Ranvier is incorporated as an ionic flux as given by Eq. \ref{eq:nbc}. Here $I_i$ is the current of each ionic concentration computed using their respective Hodgkin-Huxley ionic conductance.\\

At top and bottom boundaries;
\begin{equation}
    c_i = c_i^e, \qquad V=0 \label{eq:dbc}
\end{equation} 

At the Node of Ranvier:
\begin{equation}
    \boldsymbol{F}_i \cdot \boldsymbol{n} = f_i, \qquad f_i = \frac{I_i}{z_i F} \label{eq:nbc}
\end{equation}

where, 
\begin{equation}
    I_{Na} = \bar{G}_{Na} m^3 h (V_m-E_{Na}) \nonumber
\end{equation}
\begin{equation}
    I_{K} = \bar{G}_K n^4 (V_m-E_K) \nonumber
\end{equation}
\begin{equation}
    I_{Cl} = 0 \nonumber
\end{equation}

\subsection{Estimating effective electrical properties from ionic distributions}

\begin{figure}[h!]
  \centering
\includegraphics[width=1.0\linewidth]{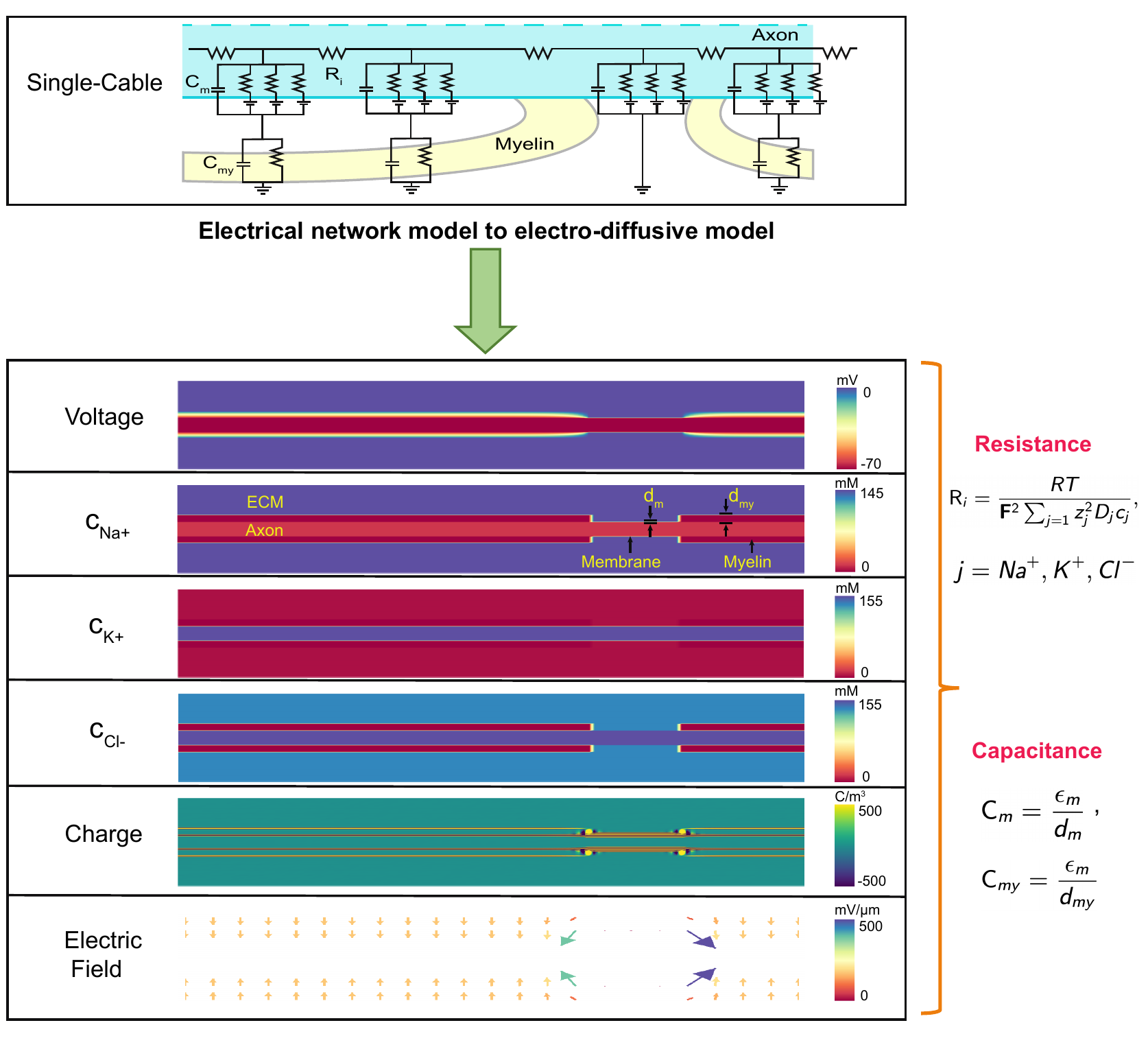}
\caption{Electrical connection between the cable theory based model and the Poisson-Nernst-Planck model. The spatial electro-diffusive PNP model comprises of multiple fields including the voltage and the concentration of the respective ions. The capacitance of the membrane and the myelin sheath are embedded in the PNP model in the form of membrane/myelin thickness. The ion channel currents are treated as a flux (Neumann boundary condition) in the spatial model. The resistance offered to the action potential along the axon can be computed using the ionic constants and the expression given in the figure. The derived fields namely, the electric field and net charge are also plotted. Note that the plot of electric field is zoomed in {around the membrane} as there is no appreciable electric field in the ECM.}
 \label{fig:electricalToSpatial}
\end{figure}

In order to compare the action potential propagation modelled using the cable theory and the PNP model, the embedded electrical properties of the PNP model should be the same as the input values of the cable theory based models. The capacitance of the PNP model can be computed using the membrane thickness and myelin thickness to obtain $C_m$ and $C_{my}$, respectively. The axial resistance offered to the conduction can be computed using the ionic constants as presented in Figure \ref{fig:electricalToSpatial}. Initially, the membrane potential is at its resting value of $-70$ mV with higher concentration of sodium ions in the extracellular region and a higher concentration of potassium ions in the intra-cellular region. The model is first equilibrated for a few timesteps until there are no fluctuations in the field variables. Next, to initiate the action potential, we assume a sodium ion influx in a region of length $5 \mu m$ near the left end of the neuronal membrane until the local potential reaches the threshold potential. Thereafter, the activation parameters regulate the depolarization/repolarization.

\section{Results}
\label{section: results}
This section seeks to establish the significance of the PNP numerical framework by analyzing the results of the three variants of the electro-diffusive model discussed in the preceding sections. First and foremost, an unmyelinated PNP model for a squid axon is investigated and distinctions with the cable theory based Hodgkin-Huxley model are probed. Subsequently, the simulation results for the PNP with myelin and PNP with myelin plus peri-axonal space model are carefully studied and a detailed comparison with the corresponding cable theory based model is presented. Neuronal axons of a rat and squid are considered for the numerical simulations. The parameters of the respective axons are collected from \cite{hodgkin1952huxley, cohen2020cell, dione2016pnp1, lopreore2008pnp2} and are listed in the Supplemental Information.

\subsection{PNP model}
\begin{figure}[h!] 
  \centering
\includegraphics[width=1.0\linewidth]{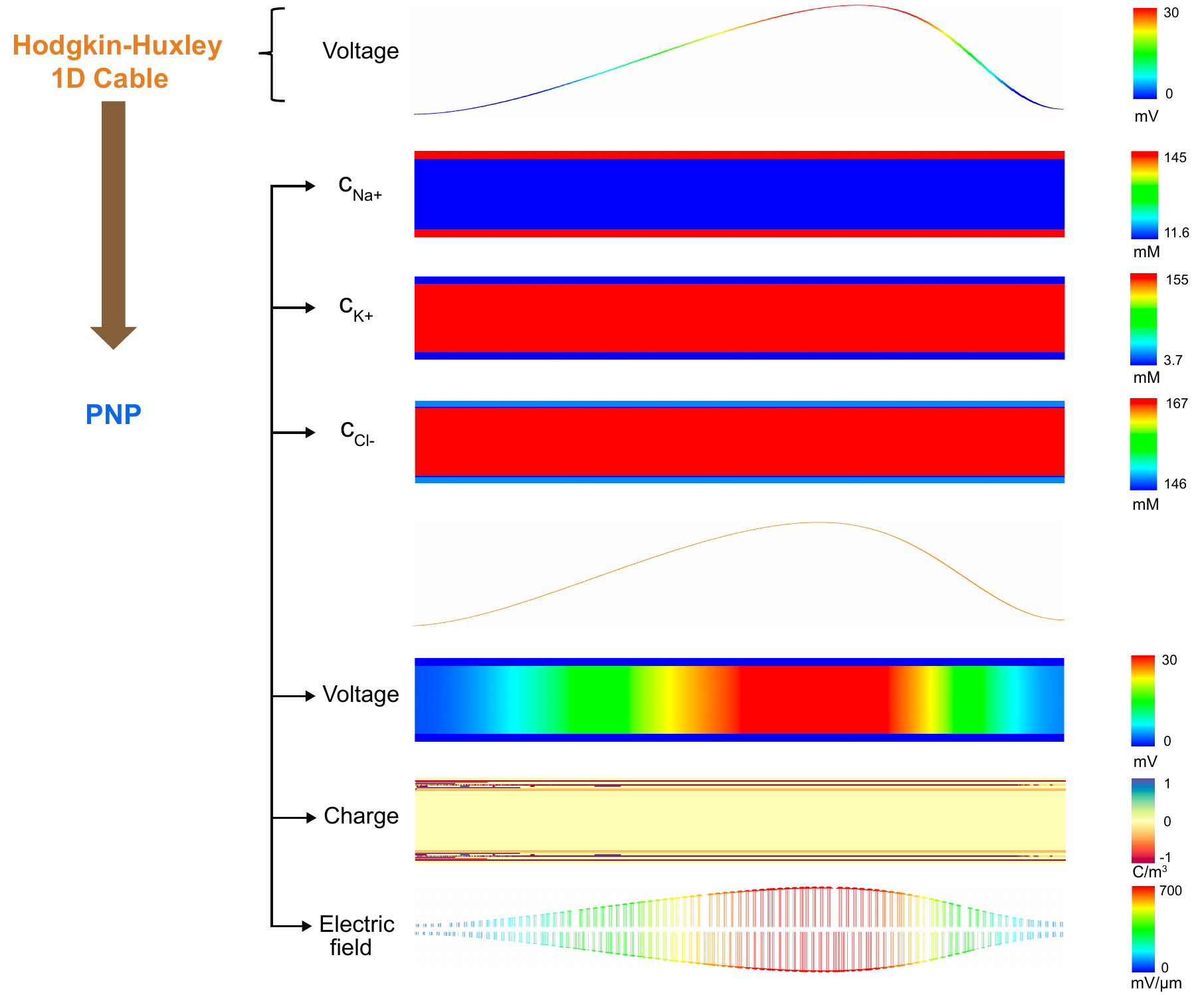}
\caption{Comparison of the cable theory based Hodgkin-Huxley to the PNP model for a squid axon. {The models have identical inputs of the membrane capacitance and resistance}. The schematics depict the capability of the PNP model to visualize the spatial distribution of the voltage, ionic concentration, net charge and the electric field. Net charge and electric field are dominant at the membrane. The profile of the one dimensional lineout of the action potential propagation from the PNP model resembles to the action potential propagation modelled using the cable theory based classical Hodgkin-Huxley model.}
 \label{fig:HHModel}
\end{figure}

Figure \ref{fig:HHModel} compares the action potential resulting from the cable theory based Hodgkin-Huxley model to the PNP model. For this simulation, the neuronal axon of a squid having a length of $20,000$ $\mu $m, a radius of $238$ $\mu$m, and a rat neuron having a length of $1500$ $\mu $m, a diameter of $1.1$ $\mu$m are considered. Note that a larger length for the squid axon is selected so that we can clearly observe the propagation profile of the action potential in the simulations. From the plots, we can clearly visualize that the one-dimensional lineout of the spatial voltage resulting from the PNP model is similar to the voltage from the cable theory based Hodgkin-Huxley model. The charge is accumulated near the membrane due to the property of the membrane to act as a capacitor. This leads to the formation of the Debye layer near to the membrane. As reflected from the plot of net charge, the variation of ionic concentration is appreciable near the membrane. The electric field is prominent throughout the membrane due to the potential difference between the intracellular and the extracellular region, however, the magnitude of this electric field varies with the propagation of the action potential along the axon. The initial equivalent resistance offered to the action potential propagation along the squid axon is computed to be {$38.18$ $\Omega$ cm} using the underlying ionic constants. The conduction velocity for the squid and the rat axon is computed to be 15.38 m/s and 0.36 m/s respectively. 

\subsection{PNP model with myelin}
\begin{figure}[h!]
  \centering
\includegraphics[width=1.0\textwidth]{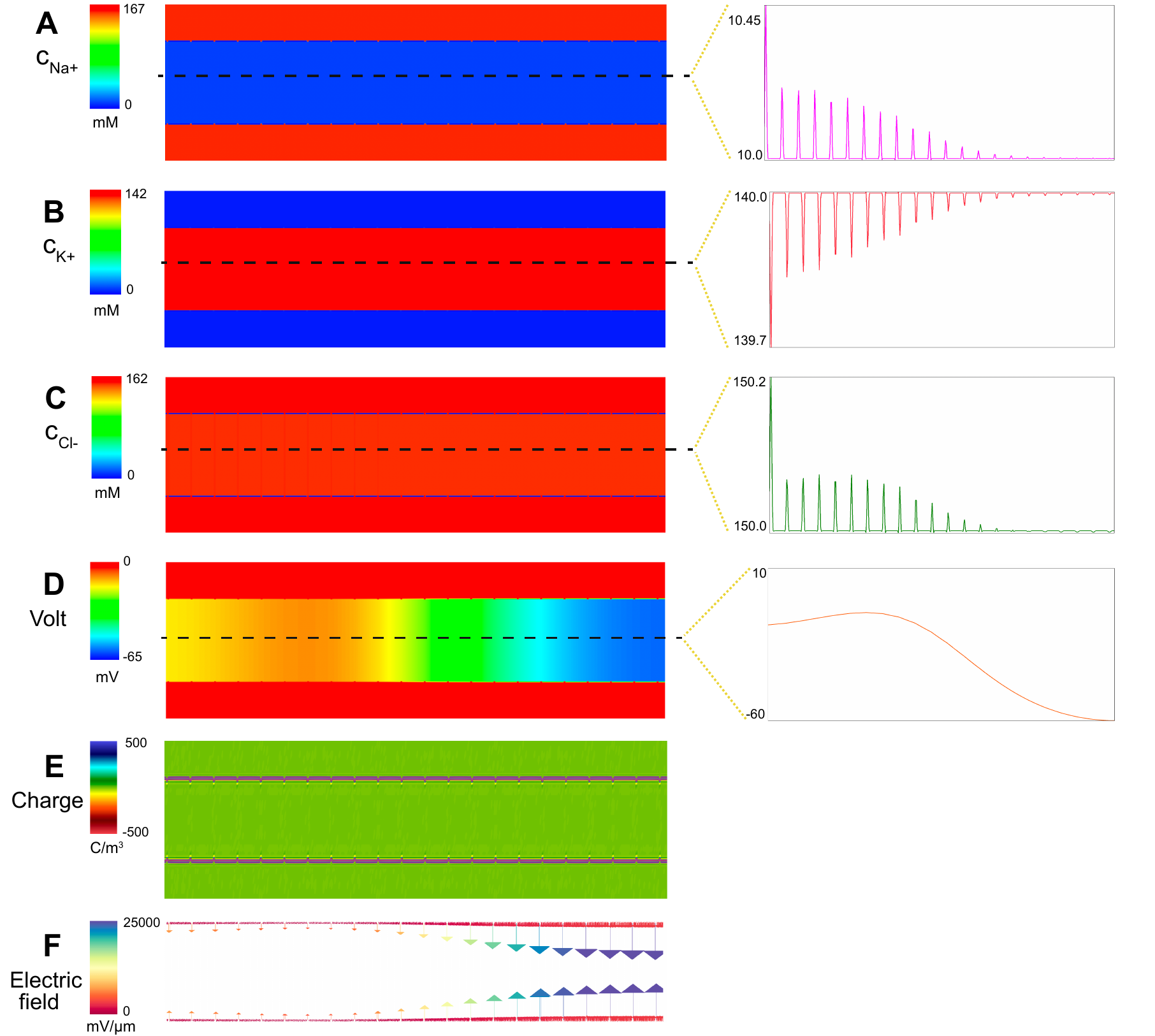}
\caption{Schematics of the various fields modelled using the PNP model with myelin for a rat axon. A lineout of the one dimensional profile of the primary fields extracted from the spatial PNP model is depicted in the right. The action potential propagates as a soliton-like wave as in the HH model. The ionic concentrations depict a saltatory conduction where it appears that the concentration jumps from one node of Ranvier to another. This is unlike the saltatory conduction observed using the one dimensional single-cable model. This is observed for all the ionic species present. Net charge is accumulated near the membrane. Electric field is dominant at the nodes of Ranvier.}
 \label{fig:SCModel}
\end{figure}

The action potential and the respective ionic concentrations modeled using the PNP with myelination model are presented in Figure \ref{fig:SCModel}. The one-dimensional lineout of the primary fields is plotted on the right. We observe that the action potential propagates as a soliton-like wave as in the single-cable model. While the saltatory conduction observed in the single-cable model is due to the potential jump from one node of Ranvier to the next, the saltatory conduction in the electro-diffusive model is observed in the respective ionic concentrations. As in the single-cable model, the saltatory conduction is due to the presence of nodes of Ranvier, and therefore the myelin sheath. Low capacitance of the myelin sheath which quickly charges and discharges, leads to fast propagation of the charge. The envelope of the one-dimensional lineout of the respective ionic concentration propagates with the action potential. We also note that there is no correlation of the myelin voltage in the electrodiffusive model as in the single cable model. The electric field, prominent at the nodes of Ranvier and the membrane is also depicted. As in the previous model, the accumulation of the net charge closer to the membrane can be observed.

Here, we discuss the mechanism of action potential propagation using the PNP electrodiffusive model. As compared to the bulk, the net charge is considerable near the membrane due to the membrane capacitance. This leads to lower net resistance in the intra-cellular region near the membrane. To initiate the action potential, there is sodium ion influx from the first node of Ranvier. The high electric field/ high voltage gradient at this node of Ranvier steers the charge to propagate at a faster velocity along the membrane as this offers a low resistance conduction path. These ions reach the adjacent node of Ranvier, resulting in the membrane potential attaining the threshold potential. At this point, the Hodgkin-Huxley activation/inactivation parameters govern the ionic flux and the propagation of action potential along the entire length of the axon. One must realize that the ionic diffusion is immensely slow but the combination of high voltage gradient/electric field at the respective nodes of Ranvier, the ionic flux through the membrane leads to the local current and thus faster propagation of the action potential along the axon. {This ability of the PNP framework to model the electric field and net charge, in addition to the action potential and the ionic distributions, make it a high-fidelity and more comprehensive model compared to the cable theory models.}

\subsection{PNP model with myelin and peri-axonal space}
\begin{figure}[h!]
  \centering
\includegraphics[width=1.0\textwidth]{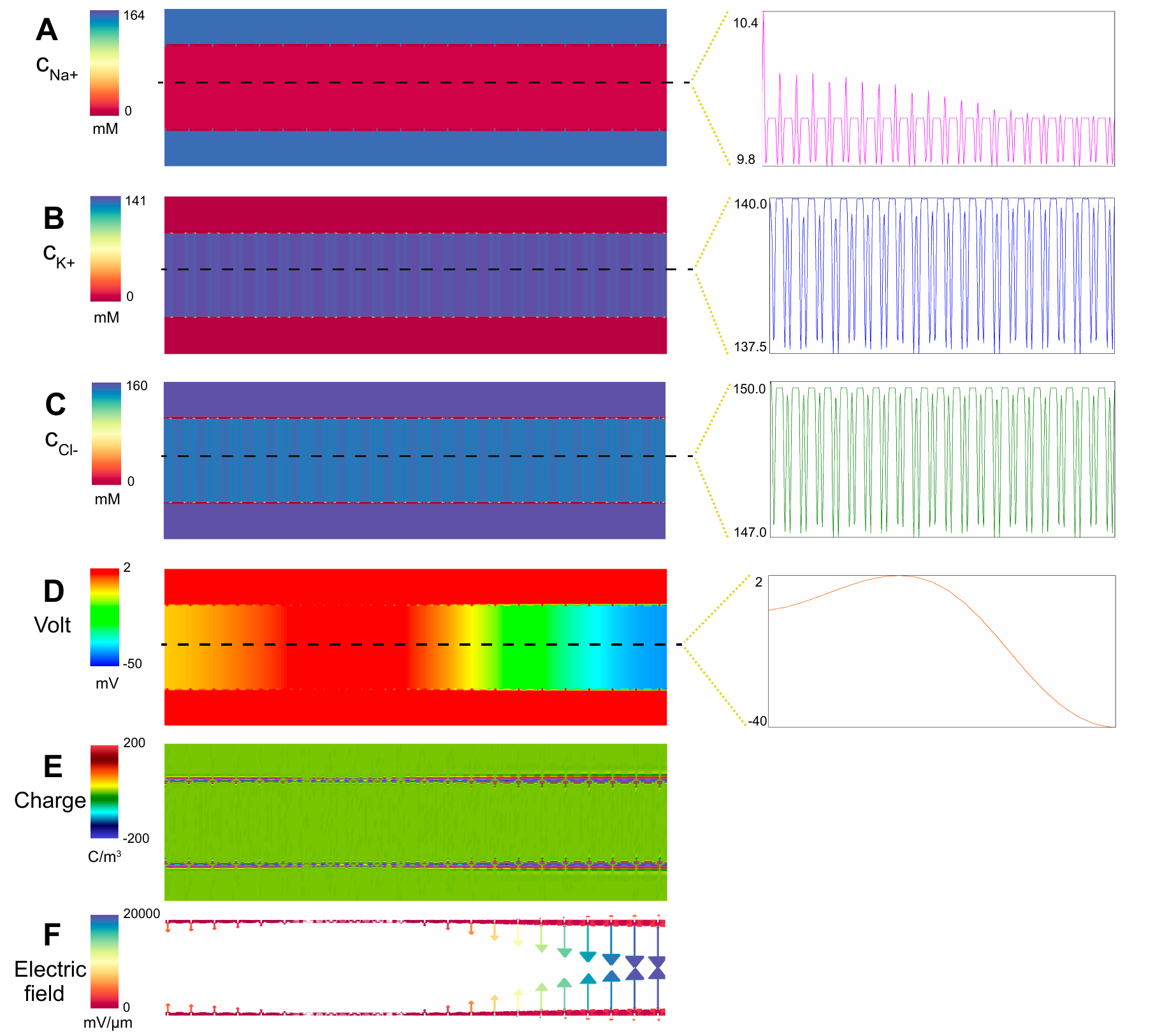}
\caption{Schematics of the fields modelled using the PNP model with myelin and periaxonal space for a rat axon. A lineout of the one dimensional profile of the primary fields extracted from the spatial PNP model is depicted in the right. The profile of the action potential is similar to earlier models. As in the PNP with myelin model, the ionic concentrations depict a saltatory conduction jumping from one node from Ranvier to the next. However, this conduction is faster here as compared to the PNP with myelin model.}
 \label{fig:DCModel}
\end{figure}

The propagation of the action potential using the PNP model with myelin and peri-axonal space is depicted in Figure \ref{fig:DCModel}. The profile of the action potential propagation is similar to the earlier presented PNP model. However, the one-dimensional lineout plot of the respective ionic concentration depicts a lot of surges due to the presence of the peri-axonal space in addition to the nodes of Ranvier. The envelope of the sodium ion concentration propagates with the action potential. However, the envelope of the potassium and the chloride ion travel relatively fast with respect to the action potential. Again, we can visualize the spatial saltatory conduction from the respective ionic concentration which jumps from one node of Ranvier to the next. As in the previous PNP model, we note the lower amplitude of action potential propagation, the dominant electric field at the nodes of Ranvier, and the accumulation of net charge near the membrane. Note that the magnitude of the electric field varies with the neuronal conduction along the axon. Simulation video of the myelinated PNP models showing the spatio-temporal evolution of all the fields is provided as part of the Supplemental Information.

\subsection{Spatial resolution of the primal fields}
\begin{figure}[h!] 
  \centering
\includegraphics[width=1.0\linewidth]{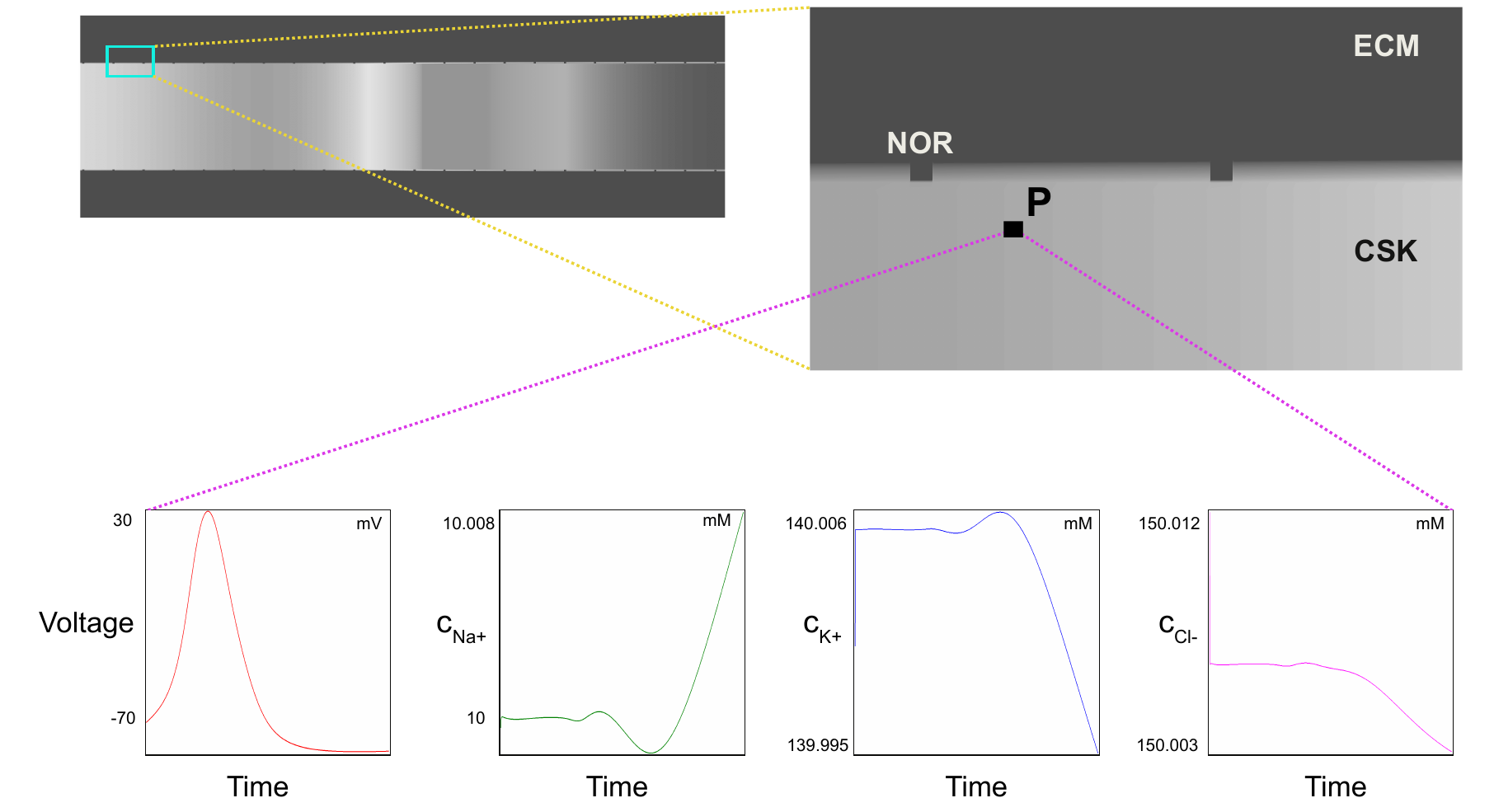}
\caption{{Spatial resolution of the primary and derived fields in the PNP with myelin model of a rat axon. The framework of the PNP based models can provide insight into the various spatial fields present at a material point on the neuronal membrane or its vicinity. The time evolution of various field values at a point `P' in the intra-cellular region is depicted in the plots.}}
 \label{fig:spatialModelFields}
\end{figure}

Based on the simulation results presented in the previous sections, one can appreciate the potential of the PNP framework to augment the voltage imaging experiments where the neuron is stained with a voltage-sensitive or ion-sensitive dye to carry out an analysis of the action potential propagation and its properties such as conduction velocity. Figure \ref{fig:spatialModelFields} demonstrates the capability of the PNP framework to determine the voltage fields and the respective ionic concentration to a fine resolution. {This fine resolution is important, as it's the small variations in the respective ionic concentrations that eventually result in the quintessential propagation of the action potential}. At any point P in the domain, one can visualize the temporal variation of the respective primary or derived fields. {Thus, in addition to the voltage imaging experiments, this model can potentially act as a digital twin for the patch-clamp/voltage-clamp experiments where the profile of the voltage or the current is plotted at a specific point in the neuron where the electrode is inserted.}

\section{Discussion on estimated velocity of action potential propagation}
\label{section:discussion}
\begin{figure}[h!]
\centering
\subfloat[\label{fig:cvRat}]{
\includegraphics[scale=1.0]{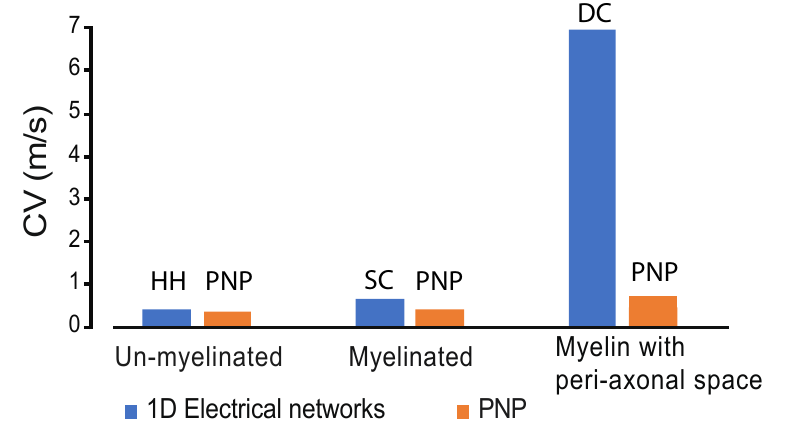}
}
\subfloat[\label{fig:cvSquid}]{
\includegraphics[scale=1.05]{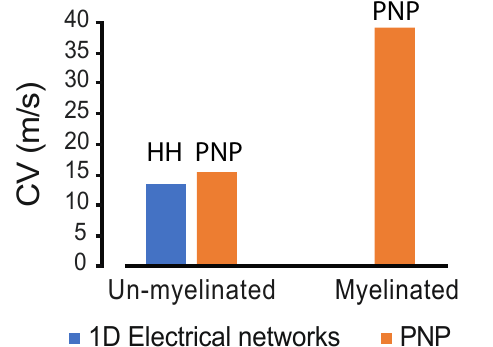}
}
\caption{Estimated conduction velocity (CV) of action potential propagation. (a) Numerical estimates of the CV of the action potential propagating through the rat neuron modeled using variants of cable theory and PNP based models. Using the cable theory based models, the CV increases due to the presence of the myelin sheath and the peri-axonal space. The same trend is observed for the PNP models but the increase in the conduction speed when the peri-axonal space is present in the PNP model, is comparatively lower than the double-cable model. (b) CV for the action potential propagating through the squid neuron modeled using cable theory and PNP models. The increase in the CV is significant when myelin is present. Here, HH, SC, and DC abbreviations are used for the Hodgkin-Huxley, Single-Cable, and Double-Cable models, respectively, to distinguish between the three 1D electrical network models.}
\label{fig:cvRatSquid}
\end{figure}

The conduction velocity for a rat axon computed using the various cable theory based models and the PNP models are compared in Figure \ref{fig:cvRat}. For each of the models based on cable theory or PNP, the conduction velocity increases due to myelination or the presence of submyelin peri-axonal space. While the conduction velocity computed using the PNP and the PNP with myelin are in close proximity to that computed from the classical Hodgkin-Huxley and the single-cable model respectively, the double cable model has nearly a nine fold increase in the speed compared to the PNP model with myelin and periaxonal space. This may be due to the fact that, while the double-cable model employs the peri-axonal resistance and the para-nodal resistance, these values for the PNP model depend on the underlying ionic constants and are therefore of the same order of magnitude as in the axonal cytoplasm. It has been assumed that the ionic diffusion coefficient is the same in the cytoskeleton of the neuronal axon and the sub-myelin peri-axonal space which may not be the case. However, once these values are obtained experimentally, we can easily incorporate this in the PNP framework. The slight difference in the conduction velocity computed using the PNP model, PNP with myelin model as compared to the cable theory based Hodgkin-Huxley, single cable model is discussed next. Initially the equivalent axial resistance of the PNP model is comparable to the cable theory based model. With the propagation of action potential and because of the nature of the membrane to act as a capacitor, the net charge is appreciable near the membrane in the intracellular region. This leads to lower net resistance near the membrane in the PNP models and the differences in the conduction velocity when compared to their corresponding cable theory based model.

For a squid, there is a two fold increase in the computed conduction velocity when the electro-diffusive model includes myelination. The conduction speed of the Hodgkin-Huxley model is close to the value computed using the PNP model. The primary difference in neuronal conduction velocity of a squid and a rat is due to the larger diameter of the squid axon. The net axial resistance in the rat axon is therefore relatively higher.

\section{Conclusion}
\label{section: conclusion}
{Cable theory based Hodgkin-Huxley, single-cable and double cable models have been extensively used to model the action potential propagation along the neuron. These models are essentially a one dimensional reduced order representation of the complex action potential dynamics, and ignore the spatial ionic diffusion which can be crucial to study propagation in non-trivial geometries and in the heterogeneous neuronal microenvironment, such as in action potential propagating through thin dimensions in dendrites, and of multiple action potentials. The Poisson-Nernst-Planck model provides a robust framework to represent the action potential propagation due to the spatio-temporal variation of the ionic concentrations at a much finer resolution. Also, the derived fields, such as the electric field and the net charge, can be easily obtained and analyzed. Moreover, as shown in this work, this model is able to emulate the various variants of the cable theory based models. This framework can also be extended to model the heterogeneous microenvironment around the neuronal membrane, such as the neuron-glia interactions.}

{To the best of our knowledge, the framework presented here for numerically estimating the conduction velocity of the spatial action potential propagating using the PNP model,  is a first-of-its-kind. Further, the spatio-temporal manifestation of the well-known saltatory conduction mechanism has been demonstrated. This electro-diffusive PNP framework can augment experiments to yield an elevated understanding of the action potential propagation, and thus assisting us to get a better insight into the functioning of our neurophysiology, and into the action potential related underpinnings of various neurological diseases.}

\section*{Acknowledgement}
\label{section:acknowledgement}
The authors thank the U.S. Office of Naval Research (ONR) for supporting this work under the ``PANTHER'' program (award number: N00014-21-1-2918) through Dr. Timothy Bentley. The authors would also like to thank Prof. Christian Franck at the University of Wisconsin-Madison and Prof. Ashfaq Adnan at the University of Texas at Arlington for insightful discussions on the subject.


\section*{References}
  
\bibliographystyle{elsarticle-num-names}
\bibliography{references}
\end{document}


\date{}
\begin{flushleft}
{\Large
\textbf\newline{Supplemental Information for ``Spatio-temporal modeling of saltatory conduction in neurons using Poisson-Nernst-Planck treatment and estimation of conduction velocity''}
}
\vspace{0.1in} 
\\
Rahul Gulati, 
Shiva Rudraraju\textsuperscript{*}
\\ 
\bigskip
Department of Mechanical Engineering, University of Wisconsin-Madison, Madison, WI 53706, USA
\\
\bigskip
*Corresponding Author \\
Email: shiva.rudraraju@wisc.edu
\end{flushleft}

\section{Hodgkin-Huxley activation/inactivation parameters}
$n$, $m$ and $h$ represent the potassium channel activation, sodium channel activation, sodium channel inactivation and are dimensionless quantities between 0 and 1, relating the ionic flux of the respective ions across the neuronal membrane. These are described by the following ODE's:
\begin{equation}
    \frac{dn}{dt}= \alpha_n (V_m) (1-n) - \beta_n (V_m) n \\
\end{equation}
\begin{equation}
    \frac{dm}{dt}= \alpha_m (V_m) (1-m) - \beta_m (V_m) m \\
\end{equation}
\begin{equation}
      \frac{dh}{dt}= \alpha_h (V_m) (1-h) - \beta_h (V_m) h \\
\end{equation}
The rate constants $\alpha_i$, $\beta_i$ are given by
\begin{equation}
    \alpha_n (V) = \frac{0.01(10-V)}{\exp(\frac{10-V}{10})-1}, \quad \quad \beta_n (V) = 0.125 \exp(\frac{-V}{80}) \\
\end{equation}

\begin{equation}
        \alpha_m (V) = \frac{2.5-0.1V}{\exp(\frac{25-V}{10})-1}, \quad \quad  \beta_m (V) = 4 \exp(\frac{-V}{18}) \\
\end{equation}

\begin{equation}
        \alpha_h (V) = 0.07 \exp(\frac{-V}{20}), \quad \quad \beta_h (V) = \frac{1}{\exp(\frac{30-V}{10})+1} \\
\end{equation}

\section{Leak current}
The leak current controls the hyperpolarization phase of the action potential propagation and therefore ensures that the membrane potential comes back to its resting state after the signal has propagated. The following expression for the leak current is adopted \cite{boucher2012leak104b}. Note that this leak is active only at the nodes of Ranvier when the myelin sheaths are present.

\begin{equation}
    I_K^{leak} = g_K^{leak} (V-E_K)
\end{equation}

\begin{equation}
    I_{Na}^{leak} = g_{Na}^{leak} (V-E_{Na})
\end{equation}

\section{Ionic pumps}
The expression for the ionic pumps embedded in the neuronal membrane is mentioned below and is adapted from \cite{lauger1991ionPump, kager2000pump104, Yu2012pump104d}. The role of the ionic pumps is to replenish the respective ionic concentration in the intracellular region to its resting state concentration after the propagation of the action potential. In some simulations, ionic pump term is ignored for better visualization of the respective fields.  
\begin{equation}
    I_K^{pump} = -2I_{max} A_{pump} \nonumber
\end{equation}

\begin{equation}
    I_{Na}^{pump} = 3I_{max} A_{pump} \nonumber
\end{equation}

\begin{equation}
    A_{pump} = \Big(1+\frac{k_K}{c_{K}^e}\Big)^{-2} \Big(1+\frac{k_{Na}}{c_{Na}^{i}}\Big)^{-3} \nonumber
\end{equation}

Here $k_i$ are the Michaelis-Menten dissociation constants for the respective ion, $c_K^e$ is the local concentration of potassium in the extracellular region and $c_{Na}^i$ is the local concentration of sodium ions in the intracellular region. 


\section{List of videos}
\begin{table}[h!]
\centering 
\begin{tabular}{c c} 
\hline 
 Filename & Description \\ [0.5ex] 
\hline 
PNP-Myelin.mp4 & PNP with myelin model for a rat axon  \\ [1.5ex]
PNP-Myelin-Peri-axonal space.mp4 & PNP with myelin and peri-axonal space model for a rat axon  \\ [1.5ex]

\hline 
\end{tabular}
\caption{List of simulation videos provided with the SI} 
\label{table:videos} 
\end{table}

\section{Table of Parameters}

 \begin{table}[h!]
\centering 
\begin{tabular}{c c c} 
\hline 
 Parameter & Value & Description \\ [0.5ex] 
\hline 
R &	8.31454 $J \cdot mole^{-1} \cdot K^{-1}$ &	Gas constant	\\ [1.5ex]
T &	279.45 $K$ &	Temperature	\\ [1.5ex]
F &	96485 $C \cdot mole^{-1}$ &	Faraday constant	\\ [1.5ex]
$D_{Na}$ &	1.33 $\mu m^2 \cdot ms^{-1}$ &	Diffusion coefficient of sodium ion	\\ [1.5ex]
$D_{K}$ &	1.96 $\mu m^2 \cdot ms^{-1}$ &	Diffusion coefficient of potassium ion	\\ [1.5ex]
$D_{Cl}$ &	2.0 $\mu m^2 \cdot ms^{-1}$ &	Diffusion coefficient of chloride ion	\\ [1.5ex]
$\epsilon_o$ & 8.88541 $e^{-12} C \cdot m^{-1} \cdot V^{-1}$ & Electric permittivity in vacuum \\ [1.5ex]
$\epsilon_w$ & 80.0 & Relative dielectric permittivity of water \\ [1.5ex]
$g_{Na}^{leak}$ &	0.065 $mS/cm^2$ &	leak conductance of $Na^+$	\\ [1.5ex]
$g_{K}^{leak}$ &	0.435 $mS/cm^2$ &	leak conductance of $K^+$	\\ [1.5ex]
\hline 
\end{tabular}
\caption{Constants used in the electro-diffusive model simulations} 
\label{table:constantsPNP} 
\end{table}

\begin{table}[h!]
\centering 
\begin{tabular}{c c c} 
\hline 
 Parameter & Value & Description \\ [0.5ex] 
\hline 
$c_{Na}^i$ &	12 $mM$ &	Initial intracellular concentration of $Na^+$	\\ [1.5ex]
$c_{K}^i$ &	155 $mM$ &	Initial intracellular concentration of $K^+$	\\ [1.5ex]
$c_{Cl}^i$ &	slightly exceeds 167 $mM$ &	Initial intracellular concentration of $Cl^-$	\\ [1.5ex]
$c_{Na}^e$ &	145 $mM$ &	Initial extracellular concentration of $Na^+$	\\ [1.5ex]
$c_{K}^e$ &	4 $mM$ &	Initial extracellular concentration of $K^+$	\\ [1.5ex]
$c_{Cl}^e$ &	149 $mM$ &	Initial extracellular concentration of $Cl^-$	\\ [1.5ex]
$r$ & 238 $\mu m$ & Radius of squid axon \\ [1.5ex]
\hline 
\end{tabular}
\caption{Neuronal parameters used for the squid axon \cite{dione2016pnp1,lopreore2008pnp2}} 
\label{table:squidParameters} 
\end{table}

\begin{table}[h!]
\centering 
\begin{tabular}{c c c} 
\hline 
 Parameter & Value & Description \\ [0.5ex] 
\hline 
$c_{Na}^i$ &	10 $mM$ &	Initial intracellular concentration of $Na^+$	\\ [1.5ex]
$c_{K}^i$ &	140 $mM$ &	Initial intracellular concentration of $K^+$	\\ [1.5ex]
$c_{Cl}^i$ &	exceeds 150.0 $mM$ &	Initial intracellular concentration of $Cl^-$	\\ [1.5ex]
$c_{Na}^e$ &	155 $mM$ &	Initial extracellular concentration of $Na^+$	\\ [1.5ex]
$c_{K}^e$ &	3.5 $mM$ &	Initial extracellular concentration of $K^+$	\\ [1.5ex]
$c_{Cl}^e$ &	158.5 $mM$ &	Initial extracellular concentration of $Cl^-$	\\ [1.5ex]
r & 0.55 $\mu m$ & Radius of rat axon \\ [1.5ex]
\hline 
\end{tabular}
\caption{Neuronal parameters used to model the rat axon \cite{muller2000ratIon, cohen2020cell}} 
\label{table:ratParameters} 
\end{table}
 
\begin{table}[h!]
\centering 
\begin{tabular}{c c c} 
\hline 
 Parameter & Value & Description \\ [0.5ex] 
\hline 
$C_m$ & 1.0 $\mu F/cm^2$ & Membrane capacitance \\ [1.5ex]
$R_i$ & 38.18 $\Omega cm$ & Computed axial resistance along the axon \\ [1.5ex]
\hline 
\end{tabular}
\caption{Squid axon values used for the PNP model to generate Figure 4} 
\label{table:PNPParameters}  
\end{table} 

\begin{table}[h!]
\centering 
\begin{tabular}{c c c} 
\hline 
 Parameter & Value & Description \\ [0.5ex] 
\hline 
$C_m$ & 1.45 $\mu F/cm^2$ & Membrane capacitance \\ [1.5ex]
$C_{my}$ & 0.166 $\mu F/cm^2$ & Myelin capacitance \\ [1.5ex]
$l_I$ & 70 $\mu m$ & Internodal distance \\ [1.5ex]
$l_{NOR}$ & 5.0 $\mu m$ & Length of node of Ranvier \\ [1.5ex]
$l_A$ & 1500 $\mu m$ & Total length of the neuronal axon \\ [1.5ex]
\hline 
\end{tabular}
\caption{Rat axon values used for the PNP with myelin model to generate Figure 5, 7 \cite{cohen2020cell}} 
\label{table:PNP-SCParameters} 
\end{table}

\begin{table}[h!]
\centering 
\begin{tabular}{c c c} 
\hline 
 Parameter & Value & Description \\ [0.5ex] 
\hline 
$C_m$ & 1.45 $\mu F/cm^2$ & Membrane capacitance \\ [1.5ex]
$C_{my}$ & 0.15 $\mu F/cm^2$ & Myelin capacitance \\ [1.5ex]
$l_I$ & 70 $\mu m$ & Internodal distance \\ [1.5ex]
$l_{NOR}$ & 5.0 $\mu m$ & Length of node of Ranvier \\ [1.5ex]
$l_A$ & 1500 $\mu m$ & Total length of neuronal axon \\ [1.5ex] 
$\delta_{pa}$ & 12 $nm$ & Thickness of peri-axonal space \\ [1.5ex]
\hline 
\end{tabular}
\caption{Rat axon values used for the PNP model with myelin and peri-axonal space used to generate Figure 6} 
\label{table:PNP-DCParameters} 
\end{table}

\vspace{1in}
\afterpage{\clearpage}
\bibliography{main}